\documentstyle[11pt,fullpage,doublespace,epsf]{article}
 \DeclareMathSizes{11}{19}{13}{9}   

\makeatletter
\@addtoreset{equation}{section}
\makeatother

\begin{document}

\title{The Tiger and the Sun: \\
Solar Power Plants and Wildlife Sanctuaries}

\author{Michael McGuigan\\Brookhaven National Laboratory\\Upton NY 11973\\mcguigan@bnl.gov}
\date{}
\maketitle

\begin{abstract}
We discuss separate and  integrated approaches to building scalable solar power plants and wildlife sanctuaries. Both solar power plants and wildlife sanctuaries need a lot of land. We quantify some of the requirements using various estimates of the rate of solar power production as well as the rate of adding wildlife to a sanctuary over the time range 2010-2050. We use population dynamics equations to study the evolution of solar energy and tiger populations up to and beyond 2050.
\end{abstract}

\section{Introduction}

An important problem in the 21st century is to construct sustainable power sources to replace or supplement coal, oil and gas whose power supplies will eventually decline. Leading candidates for renewable power sources are solar and wind power. How much of this type of power needs to be generated to significantly alter the worlds energy profile? In Table 1 we list existing power sources today \cite{energyWorld} \cite{energyUS}. From this table we can surmise that 3000 GW of solar power would have a significant global impact. Thus one can set as a goal:

\[
\textrm{Goal: Construction of 3000 GW of solar power}
\]

Similar goals of adding large amount of wind power have been discussed elsewhere \cite{wind}. How realistic is the construction of 3000 GW of solar power? The current worldwide deployment of photovoltaic solar power plants is about .955 GW so one needs more than a thousandfold increase
to achieve this goal \cite{wikisolar}. We will see that the vast land area that must be covered to construct 3000 GW of solar power is a significant factor in the deployment of this energy resource. The 3000 GW goal can be accomplished for example through the construction of 50 solar power plants of 60 GW each,  3000 plants of 1 GW or 30,000 plants of 100 MW. In section 4 we will discuss a particular case of a 60 GW solar power plant.

\begin{table}
\centering \caption{Data from \cite{energyWorld} \cite{energyUS} about energy production}
\label{pdtable1}
\begin{tabular}{|c|c|c|}
\hline
2004 Power Source &  US  &   World\\
\hline
Oil & 1340 GW & 5600 GW\\
Gas &  770 GW &  3500 GW\\
Coal & 770 GW &  3800 GW\\
Hydro &90 GW &   900 GW\\
Nuclear &270 GW & 900 GW\\
Other  & 110 GW & 130 GW\\
\hline
Total & 3350 GW & 14830 GW\\
\hline\end{tabular}
\end{table}

Another important problem in the the 21st century is the construction of wildlife sanctuaries. These sanctuaries are seen as an essential element in the preservation of endangered species. Again a significant factor in the creation of a wildlife sanctuary is the vast amount of land area that is required. Additional requirements are adequate prey and predator densities within the sanctuary. In this paper we take as an example the construction of a tiger sanctuary. Other types of sanctuaries can be discussed. However tiger sanctuaries have been shown to generates a benefit a large diversity of species \cite{Raman} \cite{Myers} \cite{Johnsingh}. How large a tiger sanctuary needs to be constructed to a have a global impact? Table 2 lists existing and past wild tiger populations by subspecies \cite{WikiTiger}. Clearly the addition of 3000 wild tigers to a wildlife sanctuary would have a major global impact, thus we can set as a goal:

\[
\textrm{Goal: Construction of a 3000 tiger wildlife sanctuary}
\]

Similar goals of adding different species can also be analyzed. How realistic is the construction of a 3000 tiger wildlife sanctuary? If one has in mind introducing a tiger population into the wild from a captive breeding program say from a national zoo as opposed from simply relocating an existing wild population \cite{relocate} this goal is extremely ambitious at least for the first generation of tigers. Besides the vast land area required, as a single tiger has a territory covering on average 10 square miles, one has the difficult issue of teaching a captive bred tiger how to hunt and survive in a sanctuary \cite{Morelli}. A project exists to reintroduce captive bred South China tiger into the wild \cite{chinatiger}. It has been documented in this project that five captive bred tigers have been trained to hunt a variety of wild prey through a learning process similar to what a mother tiger would do for her cubs. So here again about a thousandfold increase would be necessary to achieve the stated goal.

\begin{table}
\centering \caption{Data from \cite{WikiTiger} about subspecies wild tiger populations}
\label{pdtable2}
\begin{tabular}{|c|c|c|c|}
\hline
Subspecies & Wild tiger population  &    Area required & Prey population\\
\hline
Bengal & $1411 \pm  246$ & 14,000 sq. miles & 700,000\\
Indochinese  & $1300 \pm 200$  &  13,000 sq. miles & 650,000\\
Sumatran & $ 450 \pm 50 $ & 4,500 sq. miles & 225,000\\
Amur (Siberian) & $350 \pm 20$ &  17,500 sq. miles  & 175,000\\
South China & $25 \pm 5$& 250 sq. miles & 12,500\\
Javan  & $0$ &   Extinct (1980s) &  \\
Caspian & $0$ &  Extinct (1970s)&  \\
Bali & $0$ & Extinct (1940s) & \\
\hline
Total & $3,536$ & 49,250 sq. miles & 1,762,500\\ 
\hline\end{tabular}
\end{table}

This paper is organized as follows. In section 2 we discuss the rate of deployment that is possible for solar power and the land area that is required. In section 3 we discuss the rate of deployment that is necessary for the construction of a large tiger sanctuary and the land that is required. In section 4 we discuss an integrated approach that allows for the deployment of  a solar power plant as well as the construction of  wildlife sanctuary. In section 5 we discuss the main conclusions of the paper. For both  solar plants and tiger sanctuaries we used coupled logistic and Lotka-Volterra equations to discuss their growth under the influence of competition.

\section{Solar power plants}

Why build large solar power plants? There are several motivations. One of the most compelling is the desire to achieve energy independence when a country experiences declining domestic oil, gas and coal reserves.  The dependence on imported oil may threaten national security if the the supply is curtailed through world events. The decline in these type of nonrenewable reserves is often described in terms of Hubbert's peak theory \cite{Hubbert} \cite{Campbell}. The Historical Hubbert's peak data and projections are shown in Figure 1. More refined analysis with multiple peaks are contained in \cite{Berg} and \cite{Patzek}. Generalized Hubbert peaks have been found for many mineral reserves in \cite{Roper} as well as for renewable resources in \cite{Bardi} \cite{Bardi2}. Oil, gas and coal analysis has recently been done by Dave Rutledge using Hubbert's theory \cite{Rutledge}. One of his conclusions is that:

\begin{quote}
``The half-way point for ultimate oil, gas, and coal production is reached in 2019, this makes it appropriate to continue the current high growth rates for alternatives, independently of climate considerations.''

Dave Rutledge (2008)
\end{quote}

In contrast with eventual declining power sources in oil, gas and coal, one expects that alternative energy resources are poised for marked increases. In particular in the solar grand plan 
one envisions a three phase increase in solar power generation  using a combination of photovoltaic (PV) and concentrated solar power (CSP) plants \cite{Fth1}. One can also consider a wind grand plan or a geothermal grand plan. Which plan that is adopted depends on the local resources of a given region. More details related to the solar grand plan can be found in \cite{Fth2} \cite{Smesstad} \cite{Denholm1} \cite{Denholm2} \cite{Choudhary} \cite{Reynolds} \cite{Salameh}.

\begin{figure}[htbp]

   \centerline{\hbox{
   \epsfxsize=4.0in
   \epsffile{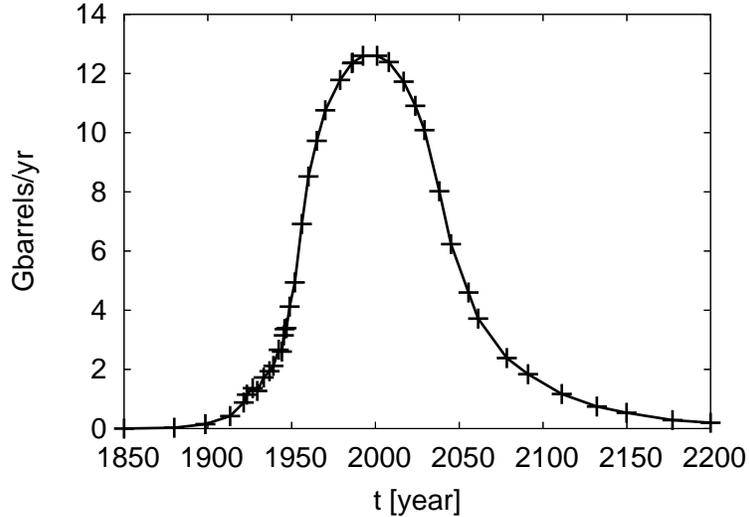}
     }
  }
 \caption{Hubbert's data  for oil depletion from 1850-2200.}

  \label{fig1}

\end{figure}

Taking the solar grand plan for inspiration one can construct a three phase increase in solar power production that can achieve the goal of adding 3000GW of solar power to the world's energy profile. In Figure 2 we plot the power capability as a function of time from this three phase plan. During the projected period of declining production of oil one sees a rapid increase in the production of solar power. Space requirements for a solar power plant vary widely depending on the design and efficiency. For definiteness consider a hypothetical 40 MW solar power plant that takes up 256 acres and produces enough power for 14000 homes. If one keeps everything in proportion then if one scales the power to 100MW the solar power plant will take up 640 acres or 1 square mile and  provide enough power for 35,000 homes. A 1GW solar power plant will take up 10 square miles and provide enough power for 350,000 homes. One can continue with this process arriving at  3000 GW of solar power that will take up a whopping 30,000 square miles and provide enough power for 1.05 billion homes worldwide. The large space is he most astounding feature of solar power deployment. When we turn to wildlife sanctuaries we shall see that there as well, staggering amounts of land a required.

Another important feature of solar power is implicit in Figure 2 is scalability of production. Scalability means that if you double resources going into production one doubles the power output after deployment. This not necessarily the case. For example there might be a single component required by a power source that can only be provided by one manufacturer in the world. In that case production must proceed serially or one after the other through this production step. This strongly limits the amount of power that can be deployed in a given time frame. For example if a 1GW power source needs a crucial component and a single manufacturer cannot make more that five such components in a year then one is limited to 5GW per year growth for that type of power source. This is somewhat analogous to Amdahl's law \cite{Amsdahl} in parallel supercomputing where one can process compute cycles only as fast as the slowest part of a computer program and this limits scalability. Scalability is achieved in the world of supercomputing through many small subprograms and hardware that build up from the microchip to petaflops of computer power. Similar scalability occurs in the production process of photovoltaic (PV) solar power as one integrates a small component subsystem into a 1GW solar power plant. Concentrated solar power (CSP) can also achieve scalability deployed in certain geographical regions.

To illustrate the potential scalability of solar power look at the three step deployment process in Table 3. In this example one assumes that four major producers of solar power can each add to the grid 1.5 GW/year of solar power during the time period 2010-2015. For this model one also assumes that these major producers can each add 15 GW/year from 2015-2020. The last and most ambitious step  is to  22.25 GW/year deployment for each supplier from 2020-2050. We also see that the number of employees are also growing during theses periods. Here we have conservatively assumed that 500 employees are required to design, construct and deploy 100 acres of solar power panels in a year.
\begin{table}
\centering \caption{Three stage solar grand plan for 2010-2050.}
\label{pdtable3}
\begin{tabular}{|c|c|c|c|}   
\hline Country  & Solar power 2010-2015 & Area required/year & Employees\\
\hline
Japan &     1.5GW/year & 15 sq. miles & 48,000\\
US    &      1.5GW/year & 15 sq. miles &48,000\\
Germany&     1.5GW/year & 15 sq. miles &48,000\\
China &      1.5GW/year & 15 sq. miles &48,000\\
\hline
Total     &       6 GW/year & 60 sq miles & 192,000\\
\hline
Country  & Solar power 2015-2020 & Area required /year &Employees\\
\hline
Japan &     15GW/year &  150 sq. miles & 480,000\\
US    &      15GW/year & 150 sq. miles & 480,000\\
Germany&     15GW/year & 150 sq. miles & 480,000\\
China &      15GW/year &  150 sq. miles & 480,000\\
\hline
Total     &       60 GW/year & 600 sq. miles & 1,920,000\\
\hline
Country  & Solar power 2020-2050 & Area required/year & Employees\\
\hline
Japan &      22.25 GW/year & 222.5 sq. miles & 712,000\\
US    &      22.25 GW/year & 222.5 sq. miles & 712,000\\
Germany&     22.25 GW/year &  222.5 sq. miles & 712,000\\
China &      22.25 GW/year &  222.5 sq. miles & 712,000\\
\hline
Total     &       89 GW/year & 890 sq. miles & 2,848,000\\
\hline
\end{tabular}
\end{table}

The rapid deployment of solar technology defines the solar power deployed as a function of time $Solar(t)$ as the piecewise function:
$$
Solar(t) = \left\{
\begin{array}{c l}
  6(t-2009) & : 2010 \le t \le 2015 \\

  60(t-2015) + 30 & : 2015 < t  \le 2020  \\

  89 (t -2020) + 330 & : 2020 < t \le 2050

\end{array}
\right\}
$$
\begin{figure}[htbp]

   \centerline{\hbox{
   \epsfxsize=4.0in
   \epsffile{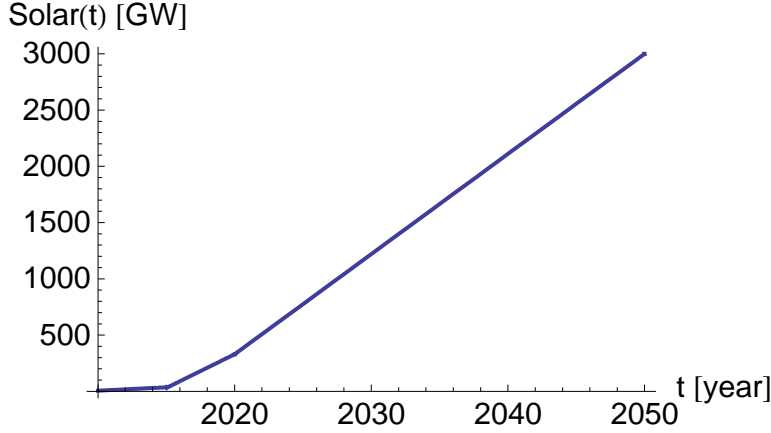}
     }
  }
 \caption{Solar power deployed as a function of time in three stage solar grand plan.
}
  \label{fig2}
\end{figure}

This function is plotted in Figure 2. The solar cumulative production is defined as the integral of the function $Solar(t)$ and is given by:
$$
ISolar(t) = \left\{
\begin{array}{c l}
  3(t-2009)^2 & : 2010 \le t \le 2015 \\

  30(t-2015)^2 +  30(t-2015) +108 & : 2015 < t  \le 2020  \\

  44.5 (t -2020)^2 + 330(t-2020) +1008 & : 2020 < t \le 2050

\end{array}
\right\}
$$

Clearly the rapid rise of the function $Solar(t)$ is very different from the decreasing profile for nonrenewables predicted by the Hubbert theory.
Also the second order polynomial cumulative production $ISolar(t)$ associated with solar deployment obey differential equations that are very different from those discussed in Hubbert's theory. For example a simple solar cumulative production $Q = 89 \frac{1}{2} \tau^2$ obeys the differential equation:
\[
\frac{d Q}{d\tau} = B \sqrt{Q}
\]
where $B = \sqrt{2(89) GW/ year}$ and $\tau = (t-2009)$. For Hubbert's theory the function on the right hand side of this equation is a quadratic function and the cumulative production and obeys:
\[
\frac{d P}{dt} = E P(1-C P )
\]
where $E$ and $C$ are constants that can be extracted from Hubbert's data and $P$ is the cumulative production associated with a nonrenewable energy source.

\begin{figure}[htbp]

   \centerline{\hbox{
   \epsfxsize=4.0in
   \epsffile{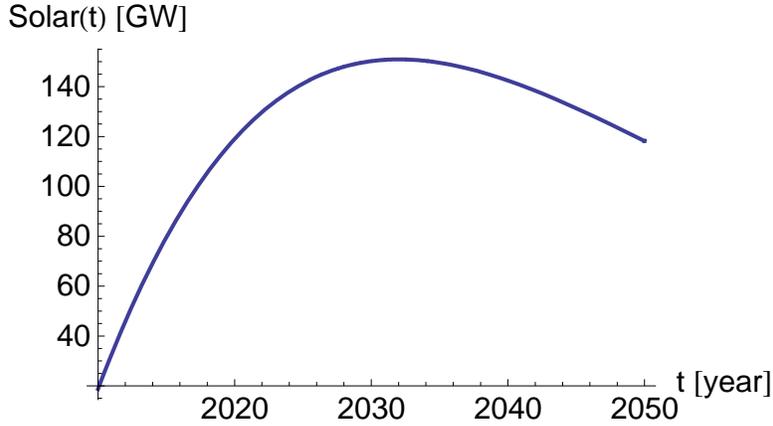}
     }
  }
 \caption{Solar energy production with a manufacturing component that is nonrenewable.}

  \label{fig3}

\end{figure}

The rapid rise of solar deployment cannot go on forever and is interesting to see what modifications to the function $Solar(t)$ and $ISolar(t)$ can take place at late times. For example if the deployment of solar cells depends on a nonrenewable resource or material that is not recycled, then as the solar cell
wears out one obtains a drop off in solar power similar to a nonrenewable resource. A simple way to generate this behavior is through the coupled differential equations
\[
\begin{array}{l}
\frac{d Q}{dt} = B \sqrt{Q}(1 - C  P)\\
\frac{d P}{dt} = E P(1-C P )\\
 \end{array}
\]
Here $Q$ is the solar power cumulative production and $P$ is the nonrenewable component. A particular solution is shown in Figure 3 for the parameters\\ $B = \sqrt{(150 GW/ year)}$,
$C = 1/561078 (GW year)^{-1}$ and $E = .0407 (year)^{-1}$. One starts with a rapid deployment of solar cells until 2030 when the dependence of the technology on the nonrenewable component becomes dominant. Thus one has to be careful in selecting a solar technology no to have implicit dependence on nonrenewable and nonrecyclable materials as this can lead to an eventual decrease in the power source. The rapid decreases can also be seen in the phenomena of extinction in population dynamics of predator prey interactions as discussed in the next section.

Another possibility for the future of the function $Solar(t)$ is that as oil reserves are depleted one still can effect the expansion of solar power. This can happen if the infrastructure for solar deployment depends on fossil fuels. For example oil burning generators used to power underground storage of solar energy. Another possibility is the the price of the fossil fuel can be adjusted down even as it is being depleted which could have the effect of lowering the urgency for which solar power plant are deployed. One can see model this effect as a form of competition between the two energy resources solar and fossil fuels. Although it is difficult to know the exact effect of lowering the price of fossil fuels has on solar development it is generally believed to have negative or adverse effect. A set of equations that could model this negative effect and include competition are:

\begin{figure}[htbp]

   \centerline{\hbox{
   \epsfxsize=4.0in
   \epsffile{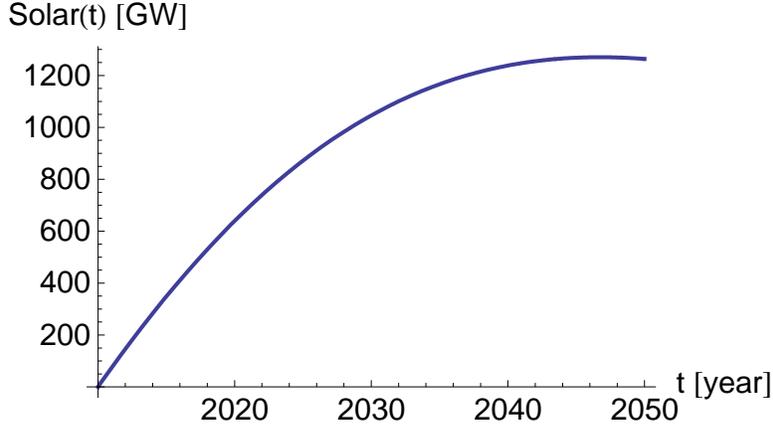}
     }
  }
 \caption{Solar energy production with the effect of competition, for example from lowering the price of oil.}

  \label{fig4}

\end{figure}

\[
\begin{array}{l}
\frac{d Q}{dt} = B \sqrt{Q} - A Q P\\
\frac{d P}{dt} =  E P(1-C P )+ A Q P\\
 \end{array}
\]
This is similar to the Lotka-Volterra equation used to describe predator/prey populations. The parameter $A$ describes the influence of low oil pricing on the solar cumulative production. A particular solution is shown in Figure 4 for the parameters $B = \sqrt{150 GW/ year}$, $C = 1/561078 (GW year)^{-1}$ , $E = .0407 (year)^{-1}$ and $A = C/(30 year)$. One sees from Figure 4 that the effect of competition with nonrenewable has lowered the deployment of solar power to less than the goal of 3000 GW by 2050. Comparing Figure 3 and 4 one sees that the effect of dependence of solar deployment on a nonrenewable resource is more severe than competition with it. To put it another way, it is better for solar to compete with nonrenewables than to team up with them. However this conclusion depends somewhat on the particular parameters that we chose to evolve for our power source dynamic equations.

\section{Wildlife sanctuaries}

Why build large wildlife sanctuaries? It is well known that certain species of wildlife are rapidly disappearing \cite{Eldredge}. A large wildlife sanctuary, if it can be made sufficiently secure, can have the effect of preserving the habitat on which this wildlife depends and thus adding to it's survival. In this paper we single out a tiger wildlife sanctuary because the tiger is severely endangered in the wild and needs a lot of space within which to define separate tiger territories. This large space provides habitats for many other species besides the tiger and it's prey. Thus the tiger  is a so called umbrella species which, if can be saved, benefits a great many other species \cite{Karanth}. Recent research has also indicated that large apex predators like tigers have addition benefits to the animals within it's territory, such as restricting the use of available habitats and food sources, and avoiding overgrazing \cite{Ale} \cite{Leopold}. Also tigers as a species are not as directly effected by global warming as say polar bears. Tiger numbers are not strongly dependent on temperature. They can hunt effectively over a wide temperature range as well as type of terrain.  An exception is tiger populations located in swampy areas which are effected by a rise in sea level \cite{Sunderbans}.  

How much area is required for a tiger sanctuary? The estimate of area we use is an average of 10 square miles per tiger. This just a rough estimate and varies from one subspecies to another based on prey density. An Amur tiger can require a much larger area for example. However one can see as indicated in the introduction adding tigers to a wild life sanctuary requires a great deal of space. Adding 3000 tigers to a Wildlife sanctuary would require an estimated 30,000 square miles of space. This is an extremely ambitious goal perhaps more so than adding 3000GW of solar power to the grid. Beside the large land area one also has to add prey to maintain an acceptable prey density. Field studies of prey density in 11 reserves in India indicate that a tiger needs 3.3 tons of prey per year to sustain itself and a ratio of 500 prey animals for every tiger \cite{Uma} \cite{Cox} \cite{Karanth1} \cite{Karanth2} \cite{Eisenberg} \cite{Johnsingh2} \cite{Schaller}. A tiger needs to kill a large prey about once a week and about ten percent of the available prey animals in it's territory in a given year. We can use the data from these field studies to determine the size of the number of tigers and prey as the sanctuary grows. Similar analysis for the Amur tiger has been done in \cite{Stephens1} \cite{Stephens2} \cite{Carroll}.

In Table 4 we list the requirements for a 3000 tiger and 300 tiger wildlife sanctuary. In this table we break down the construction of the sanctuary in three stages just as in the solar grand plan. We plot the projected tiger population for a local 300 tiger sanctuary in Figure 5 over the time range of 2010 to 2050.  
In the period from 2010-2050 we can increase the population of tigers and prey artificially by bringing them in from breeding programs. In addition the tigers must be trained to hunt. The efficiency of this training process is not known. However a low efficiency can be compensated somewhat by increasing the population from the breeding program. 

\begin{table}
\centering \caption{Near, Mid term and long range accumulated projections for ten tiger sanctuaries (World) and a single sanctuary (Local).}
\label{pdtable4}
\begin{tabular}{|c|c|c|c|}   
\hline Tigers added (World) & Area required & Prey population & Period\\
\hline
 30 &  300 sq.  miles &  15000 & 2010-2015\\
300    & 3000 sq. miles & 150,000 & 2015-2020\\
2670  & 26,700 sq. miles &  1,335,000 & 2020-2050\\
\hline
Total   3000   & 30,000 sq. miles  & 1,500,000 & 2010-2050\\
\hline
Tigers added (Local) & Area required & Prey Population & Time Period\\
\hline
 3 &  30 sq. miles & 1500 & 2010-2015\\
30    & 300 sq. miles & 15000 & 2015-2020\\
267&   2,670 sq. miles & 133,500 &  2020-2050\\
\hline
Total   300   &  3000 sq. miles & 150,000 & 2010-2050\\
\hline

\end{tabular}
\end{table}

\begin{figure}[htbp]

   \centerline{\hbox{
   \epsfxsize=4.0in
   \epsffile{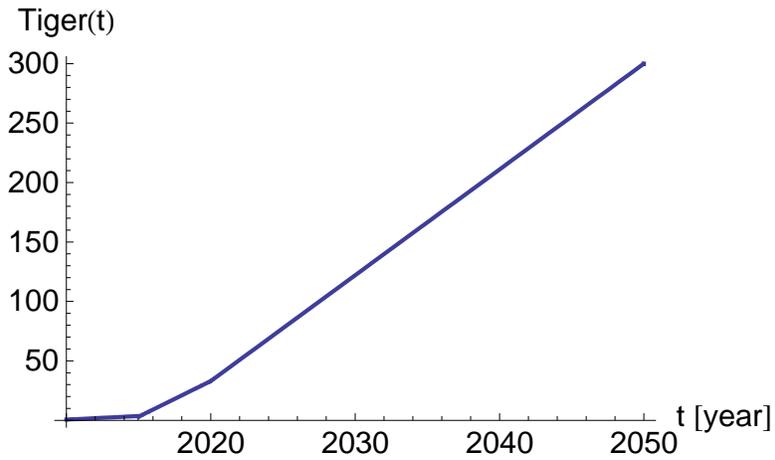}
     }
  }
 \caption{Tiger population for a local tiger sanctuary for 2010-2050. The increase is do to the transfer from a captive breeding program together with training in hunting.}

  \label{fig5}

\end{figure}

\begin{figure}[htbp]

   \centerline{\hbox{
   \epsfxsize=4.0in
   \epsffile{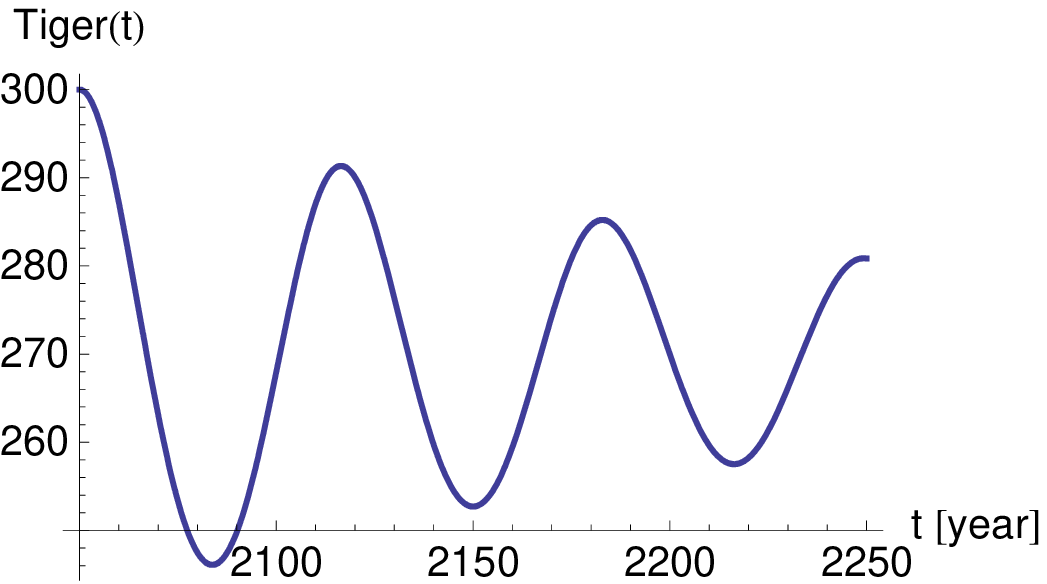}
     }
  }
 \caption{Tiger population for a local tiger sanctuary for 2050-2250.}

  \label{fig6}

\end{figure}

\begin{figure}[htbp]

   \centerline{\hbox{
   \epsfxsize=4.0in
   \epsffile{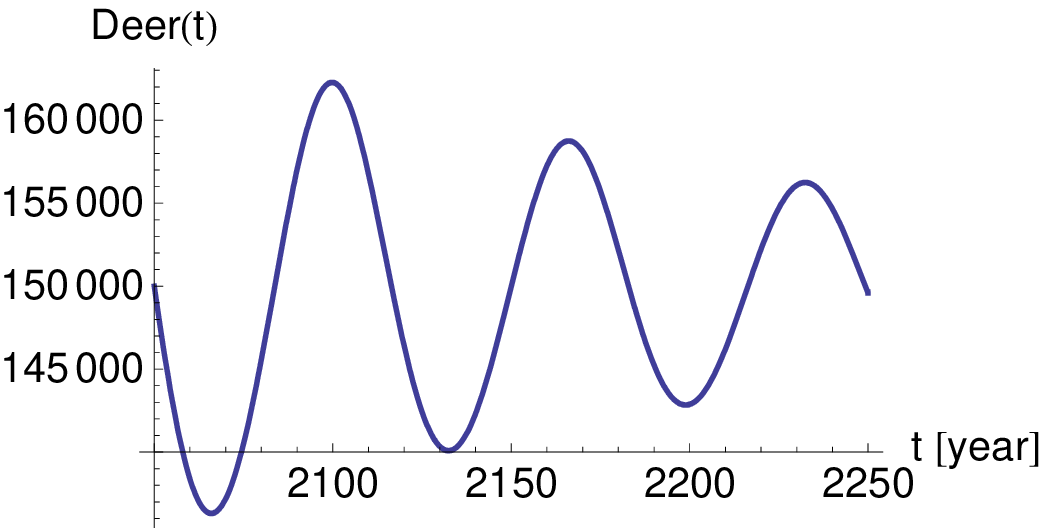}
     }
  }
 \caption{Deer population for a local tiger sanctuary for 2050-2250.}

  \label{fig7}

\end{figure}

\begin{figure}[htbp]

   \centerline{\hbox{
   \epsfxsize=3.0in
   \epsffile{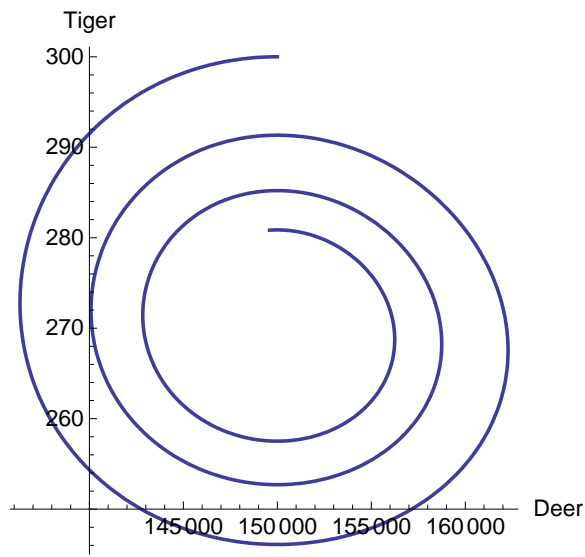}
     }
  }
 \caption{Parametric plot of tiger and deer population for 2050-2250 with time counterclockwise along the curve spiraling inwards.}

  \label{fig8}

\end{figure}

After 2050 the tiger sanctuary will be a full capacity and the population can develop similar to the existing tiger sanctuaries of today. Tiger and prey populations can be analyzed using the competitive Lotka-Volterra equations \cite{Leigh} which are:
\[
\begin{array}{l}
\frac{d Q}{dt} = B Q(1-C Q) - A Q P\\
\frac{d P}{dt} = -D P +  E P Q\\
 \end{array}
\]
Here $B$ is the birth rate of the prey (for example deer)with cumulative $Q$ and $D$ is the death rate for the tiger population whose cumulative is $P$. $A$ is a parameter which governs the rate of successful hunting by the tiger, $E$ is a parameter that describes the addition to the tiger population due to increased ability to successfully hunt, $C$ is a parameter which keeps the prey population from increasing exponentially in the absence of predators and is inversely related to the carrying capacity. 

In these equations deer are taken as representative of the prey populations. In actual preserves tiger prey vary among different animals. One can generalize the Lotka-Volterra equation to include several prey species. A large variety of behaviors can be found among the solutions to these multi-prey equations \cite{Smale}. Thus one needs to determine the parameters of the model accurately for each prey species from external field studies to ensure the stability of the population solution. Besides several prey species one can also take into account several competing predators in the equations, such as the leopard, wild dog, or jackal. Of particular interest is the equations governing the apex predator which in our case is the tiger. Recent work indicates that the existence of an apex predator adds to the stability of the competitive Lotka-Volterra equation \cite{Ohta}. Beside the Lotka-Volterra equation that governs the total populations one can also use stochastic methods that provide spatial dependence of the predator-prey systems \cite{Satulovsky}. These are typically of greater computation cost but can be run effectively on large parallel computers \cite{predatorprey} \cite{predatorprey2}.

In Figures 6-8 we plot the tiger and prey population for a local tiger sanctuary using the competitive Lotka-Volterra equations with parameters $B = .1 (year)^{-1}$, $C = 1/1,500,000$, $D = .1 (year)^{-1}$, $A = 1/3000 (year)^{-1}$ and $E = 1/1,500,000 (year)^{-1}$.
For this set of parameters the tiger and prey populations oscillate about a fixed point over many generations. The parameters were chosen so that near the fixed point the ratio of prey to tiger populations approached 500:1 and the tigers eat 10 percent of the prey population per year as was indicated in the field studies of todays tiger preserves. From the figures one can see that if the tigers can be trained to hunt as effectively as the wild tigers of today, that is if the coefficient $A$ is large enough, then a sustainable population of tigers and prey can be obtained. 

\section{Integrated approach}

\begin{figure}[htbp]

   \centerline{\hbox{
   \epsfxsize=3.0in
   \epsffile{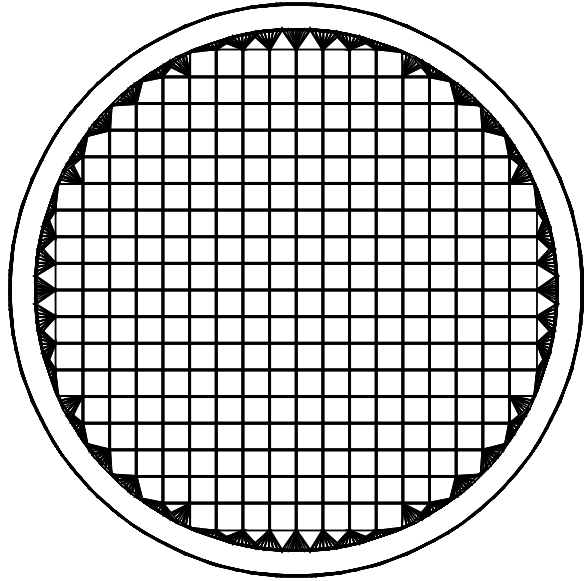}
     }
  }
 \caption{Schematic view of integrated solar power plant located in the area between the outer two circles and tiger sanctuary located in the interior of the inner circle. The radius of the inner circle is 30.9 miles and the outer circle 33.9 miles.}

  \label{fig9}

\end{figure}

Solar power plants have greater efficiency if operated in a desert, however if a country doesn't have a desert, solar power plants can also operate in temperate regions. Also if located in a desert one has potential costs in transporting the energy to more hospitable locations. In such situations one can consider an integrated approach where the wildlife sanctuary and solar power plant coexist.

One such integrated approach is indicated in Figure 9. It consists of
a wildlife sanctuary surrounded by a solar power plant. Two  outer circles bounds the solar power plant and an inner bounding circle bounds the wildlife sanctuary. One must insure that the inner circle contains a sufficient barrier to prevent the entrance of wildlife into the region of the solar power plant \cite{Walker}.  

For example, a Tiger sanctuary with 300 tigers and 3000 square miles with 150,000 prey animals would have a inner radius of 30.9 miles. This could be bound by an solar power plant in the form of an annulus with width 3 miles and area 610.35 square miles. The solar power plant would supply 61 GW of electric power. This is enough electricity to power over 20 million homes.

The cost of this integrated system can be determined using the estimates of \cite{Fth1} where a 3000 GW Solar power system was estimated to be 420 Billion dollars. Thus a 61.1 GW system would cost approximately 85.4 Billion dollars. The main assumption here is that the installed cost of solar power will be reduced from 4 dollars/Watt to about 1.2 dollars/Watt by 2050. The highest cost to date for the construction of a tiger sanctuary is 153 million over five years from the Indian government for security, population tracking and villager relocation \cite{relocate}. The cost of the wildlife sanctuary is thus only .18 percent of the cost the solar power system.

This double circle arrangement has some advantages for security and defense as well as for military tactics \cite{Caesar}. The outer circle protects the energy infrastructure while the inner circle protects the sanctuary. The area between them contains the solar power plant and the resources needed to maintain it. The integrated approach solves the security aspects of the wildlife sanctuary \cite{Panwar} \cite{Nagothu} \cite{Kenney} by coupling it to the security of the renewable energy infrastructure. Another advantage of this double circle design is that power lines and energy transport from the solar plant are attached to the outer circle only. Roads and service entrances are not requires inside the sanctuary. This allows the sanctuary portion of the facility to be contiguous allowing for migration of prey animals and greater diversity.

\section{Conclusion}

We have studied the space, economic and manpower required to achieve the goal of adding 3000 GW of solar power to an electrical grid as well the goal of adding 3000 tigers to a wildlife reserve. Due to a numerical coincidence the space required to generate 1 GW of solar power was about the same space required for a tiger, namely 10 square miles or 6400 acres. We emphasized scalability and rate of deployment in achieving both goals. 

We used modified logistic equations to study both the growth of solar power under competition and the growth of tiger populations interacting with prey. Because the equations are similar one sees that the same analysis methods can be fruitfully applied to both problems studied in this paper. Finally we introduced an integrated approach for deploying solar plants and wildlife sanctuaries within which both can coexist despite the large land areas involved.

One difficulty with wildlife conservation is that it that it is often tied to human problems. This is because most of the wildlife is located near poor rural populations as those areas have less habitat destruction. One has to make a choice as resources are limited. Ultimately one has to help the human population.

An advantage of the integrated approach is that it ties wildlife conservation to a human solution, one which must be urgently implemented. The solar energy infrastructure must surely be secured due to its strategic importance. This security extended to the integrated wildlife sanctuary within, solves on the main costs how to maintain security over a 30,000 square mile reserve. Thus in this approach one helps the wildlife as well as the human population. Indeed one is beginning to see the use solar power bordering tiger reserves as an application of electrification in rural areas \cite{Chakrabarti} \cite{Moharil}
 
\section*{Acknowledgments}
I wish to acknowledge useful discussions with Dave Rutledge and Shigemi Ohta.
This manuscript has been authored in part by Brookhaven Science Associates, LLC, under Contract No. DE-AC02-98CH10886 with the U.S. Department of Energy.


\begin{thebibliography}{100}
\bibitem{Choudhary}
S. Choudhary, ``Are solar, hydroelectric, and wind power conceivable alternatives to fossil fuel in the future?''
\end{thebibliography}
\end{document}